**Synthesis and guided assembly of niobium trisulfide nanowires and nanowire chains by chemical vapor deposition**

Thang Pham[1,3]*, Arindom Nag[3], Kate Reidy[1,#], Michael A. Filler[2]* and Frances M. Ross[1]*

1. Department of Materials Science and Engineering, Massachusetts Institute of Technology, Cambridge, MA 02139, USA
2. School of Chemical & Biomolecular Engineering, Georgia Institute of Technology, Atlanta, GA 30332, USA
3. Department of Materials Science and Engineering, Virginia Tech, Blacksburg, VA 24060, USA

# Current address: Miller Institute for Basic Research in Science, University of California, Berkeley, California 94720, USA
* Corresponding authors
Email: thangpham@vt.edu (T.P.), michael.filler@chbe.gatech.edu (M.A.F.), fmross@mit.edu (F.M.R.)

**Abstract**
One-dimensional (1D) nanostructures of transition metal trichalcogenides (TMT) show unique properties through the combination of their anisotropic bonding and low dimensionality. Scalable synthesis approaches that enable control over the morphology, dimensions, and interfaces of 1D TMTs with other nanoscale materials could allow these properties to be used in novel devices. Here, we report chemical vapor deposition of a 1D TMT, namely niobium trisulfide ($NbS_3$) in the form of nanowires, on different substrates, including bulk substrates (amorphous $SiO_2$/Si and crystalline c-sapphire) and several two-dimensional (2D) van der Waals materials (graphene, h-BN, CrSBr). We demonstrate high growth yield with axial growth rates of up to 40 μm/min and with two different growth modes: short nanowires of rectangular cross-section, and unusual long, "chained nanowires" up to 100 μm in length with sawtooth morphology. We discuss a mechanism that accounts for the two morphologies and discuss how the structure can be tuned through substrate choice and growth conditions. We further demonstrate guided assembly at the edges of graphene and h-BN, as well as epitaxial growth on few-layer CrSBr and c-sapphire. These results open pathways to explore scalable synthesis and directed assembly of 1D TMT nanomaterials in unique morphologies.

**1. Introduction**

Van der Waals (vdW) materials[1] display unique physics and chemistry due to their anisotropic bonding, combining weak vdW bonds along some crystallographic directions and strong covalent bonds along others[2]. Extensive studies of atomically thin nanostructures composed of layered (two-dimensional or 2D) vdW materials, for instance few-to-single layer graphene, hexagonal boron nitride (h-BN) or transition metal dichalcogenides (TMDs), have renewed interest in the nanostructured forms of other types of vdW materials[1,3,4]. One particular material class, the so-called "one-dimensional" (1D) vdW materials[2,5], has become prominent due to the strongly anisotropic electrical, thermal, and optical properties inherited from their 1D character[2,3,5,6].

Amongst 1D vdW materials, transition metal trichalcogenides (TMTs) with a general formula of $MX_3$, where M is a group IV (Ti, Zr, Hf) or V transition metal (Nb, Ta, V) and X is a chalcogen (S,

Se, Te), have been studied extensively in the form of bulk crystals. The structural motifs in these materials consist of trigonal prismatic $MX_6$ units assembled into infinite parallel chains of face-sharing polyhedra. The bulk crystals show strongly correlated physics[7–9], notably a competition between superconductivity and charge density waves (CDW)[10]. At the nanoscale, sub-nanometer-thick $MX_3$ has served as a platform to examine the interplay of nanoscale confinement and interactions between the chains, illustrated by size-dependent thermal conductivity in $NbSe_3$[11] and a size-dependent superconductivity-CDW crossover in $ZrTe_3$[12]. In addition to size-dependent properties, some $MX_3$ such as $NbS_3$ have shown stacking-dependent electronic properties[13], ranging from metallic in $NbS_3$ phase-II, III and V to semiconducting in $NbS_3$ phase-I and IV. $NbS_3$ also hosts unusual properties, such as three CDW transitions[14] and field-induced sliding of CDWs[15] (a phenomenon in which above a critical field, the CDW de-pins from the underlying lattice or defects, giving rise to its non-linear response). Creating a low-contact resistance interface between a nanostructured $MX_3$ and other nanoscale materials, such as 2D materials, promises even more functionalities and potential applications, for example in catalysis[16,17] and optoelectronics[18,19].

Motivated by the above properties and potential applications, there have been several attempts to obtain nanoscale structures of $MX_3$. The most common method is a top-down approach, where a bulk single crystal of $MX_3$ is grown by chemical vapor transport (CVT) and nanoscale structures are obtained by either mechanical or chemical exfoliation[3,20]. This approach works for a range of $MX_3$ compositions[3,9], but does not yield a high level of control over the sizes of the nanostructures or the interface between $MX_3$ and its support materials[2]. In contrast, bottom-up methods are those such as chemical vapor deposition (CVD) and vapor-liquid-solid (VLS)[21] growth in which nanostructures are formed directly on the support. We have recently reported that the VLS method can create $MX_3$ nanowires using Au as the catalyst, but only if salt (NaCl) is also present to assist growth[22]. The salt enables growth over a wider process window, resulting in a greater extent of control over $MX_3$ nanowire structures and compositions compared to top-down synthesis. The CVD method is simpler in concept since a metal catalyst is not required. It has been used to form mono- and dichalcogenide nanostructures[23], but with limited reports regarding trichalcogenide nanomaterials[3].

Here, we demonstrate CVD as a bottom-up and scalable method to synthesize distinct types of $MX_3$ 1D nanostructures. We exemplify this by growth of $NbS_3$ nanowires, achieving growth rates up to 40 μm/minute at temperatures of 650-750°C. We show that synthesis on an amorphous $SiO_2$/Si substrate results in two modes for nanowire growth. One mode consists of separated and relatively short $NbS_3$ nanowires, while in the other mode the nanowires assemble via end-to-end connections to form unusual segmented "chained nanowires" tens to hundreds of micrometers in length. We show that the nature of the substrate can yield morphologies that also include nanowire arrays at the edges of graphene and h-BN flakes and epitaxial nanowires on a 2D substrate with in-plane anisotropy (CrSBr) and on a crystalline 3D substrate (c-sapphire). We discuss a nucleation and diffusion-limited mechanism that accounts for these distinct morphologies based on the synthesis parameters, the time-dependence of nanowire dimensions, and the atomic level structure of the nanostructures. Lastly, we discuss the generality and application of these diverse growth morphologies.

## 2. Results and Discussion

## 2.1 Nanowire morphologies

Figure 1 shows an overview of different 1D nanostructures obtained using the procedure described in *Experimental Section* and illustrated in Fig. SI1. The NbS$_3$ is grown via salt-assisted chemical vapor deposition (CVD) using powder precursors (Nb and S) and an alkaline metal halide (NaCl) as a growth promoter[24,25]. The important synthesis parameters are summarized in Fig. SI1a, including the growth temperature (650-750 °C), growth time (1-45 minutes), flow rates of Ar carrier gas (50-150 sccm) and the substrate choice and preparation. We observe two morphologies of the 1D nanostructures: (i) separated, short nanowires (Figs. 1a-c), which we refer to as Mode 1, and (ii) long nanowires (Figs. 1d-g), which we refer to as Mode 2. These long nanowires show an unusual morphology in which nanowire segments are connected over lengths of tens to hundreds of micrometers, show sawtooth surface structure and can cover cm-sized areas of the substrate (Figs. SI1c-e).

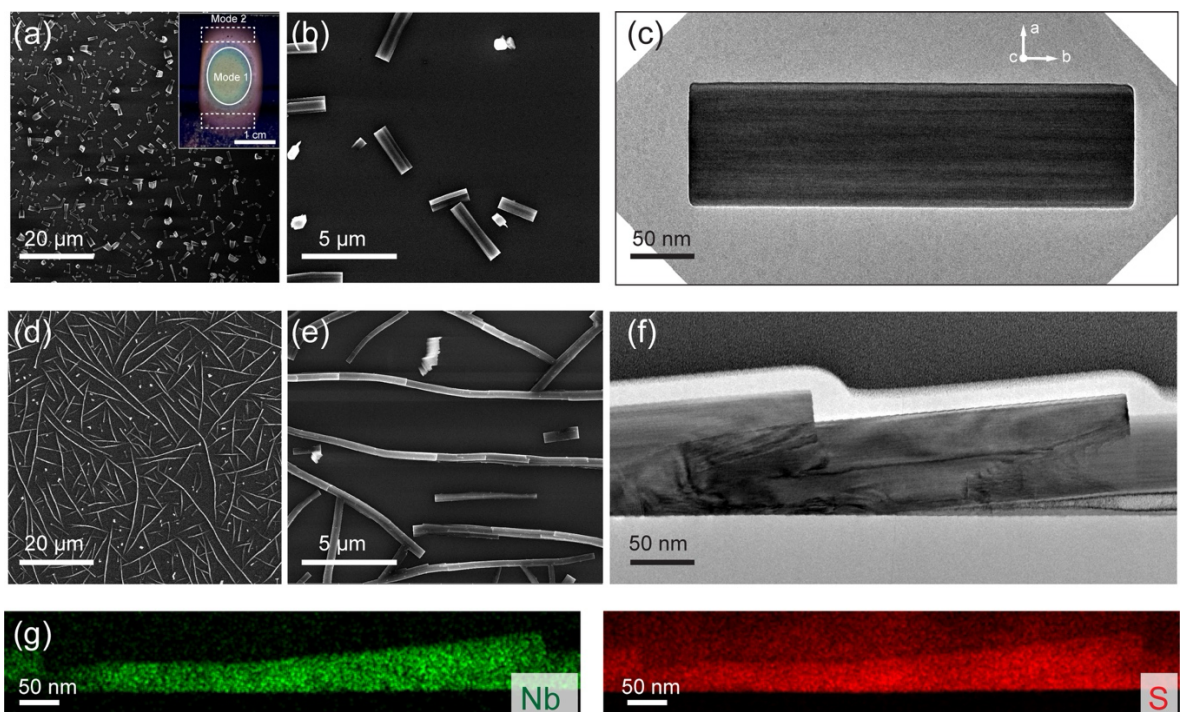

**Figure 1.** CVD growth of NbS$_3$ nanowires. (a-c) Mode-1 separated nanowires. In (a-b), the growth was on SiO$_2$/Si substrate, while in (c) the substrate is suspended graphene. The long axis of the nanowire is in the b direction. The inset of (a) shows a photograph of a post-growth substrate where the areas having Mode-1 and Mode-2 nanowires are indicated. (d-f) Mode-2 long nanowires induced by cooperative growth. (g) STEM-EDS elemental mapping displaying Nb and S distribution in a cross-sectional sample of a Mode-2 nanowire. Note that the signal in S mapping outside the nanowire area is most likely an artifact from the tail of the C peak arising from the carbon coating layer used in FIB sample preparation.

Mode 1 nanowires are shown in more detail in Figs. 1a-c. This morphology is usually found at the center of the substrate that directly faces the metal precursor (indicated in the inset of Fig. 1a and Fig. SI1b) where the precursor fluxes reaching the substrate are presumably highest. The SEM images in Figs. 1a-b and TEM image in Fig. 1c yield an average width and length of 0.35 ± 0.05 µm and 3.6 ± 0.9 µm, respectively, for a growth time of 45 min. In scanning transmission electron microscopy (STEM), measurements show lattice parameters a = 3.36 ± 0.14 nm and b = 9.78 ± 0.22 nm (representative images with a diffraction pattern are shown in Figs. SI2a-b). In combination with chemical identification and mapping by STEM-EDS (Figs. SI2c-d), the microscopy characterization indicates that the nanowires are $NbS_3$ phase-II, a metallic phase of $NbS_3$ [13,26].

Mode 2 nanowires are shown in Figs. 1d-g. This morphology is formed with identical parameters to those that produce Mode 1 nanowires, except that they are found at the two ends of the growth area (dotted rectangles in the inset of Fig. 1a). The structures are remarkably long, 70.4 ± 18.1 µm for the same 45-min growth, while their width, 0.35 ± 0.07 µm, is comparable to Mode 1 nanowires. A closer look in SEM (Fig. 1e), cross-sectional TEM (Fig. 1f) and EDS elemental mapping (Fig. 1g) reveals that each structure consists of many segments, each of which appears similar, structurally and chemically, to an individual nanowire obtained in Mode 1. These are connected end-to-end to create a long segmented nanowire with a sawtooth surface. The total number of segments that make up a Mode 2 nanowire varies from 10-50. While the observation of Mode 1 nanowires is expected with the elongated (1D) morphology plausibly explained by their anisotropic bonding[13,27], the appearance of Mode 2 chained nanowires is unexpected and to the best of our knowledge, has not been reported before. We will discuss the potential growth mechanisms of these two modes below, aiming to explain the synchronized fashion of the arrangement of segments that creates the unusual structural features of the long Mode -2 $NbS_3$ chained nanowires.

**2.2 Proposed growth mechanism**

To explore the characteristics of these self-assembled structures we first start by considering the Mode 1 nanowires. In our salt-assisted CVD synthesis, alkaline metal halide (NaCl) powder is assumed to react with niobium powder at the growth temperature (650-750 °C) to create volatile intermediates (e.g., $NbOCl_x$), as suggested in the literature[24,25]. The Nb-containing fluxes are then delivered to the substrate together with the S flux by the carrier gas (Ar). At the shortest growth times, faceted islands are visible (Figs. SI3-4). As growth time increases the structures retain their faceted shape, growing in both diameter and length, and the nucleation density increases. We can infer a relatively short surface diffusion distance from the spacing of nuclei, which is below 1 µm. The fact that the structures grow with well-faceted surfaces indicates that when atoms arrive at the growing nucleus they readily diffuse along the surface before incorporating. The elongated shape suggests a preferred incorporation site at the end of the nanowire. We assume these atoms can diffuse rapidly along the nanowire length and attach to the ends of the covalent atomic chains that comprise the structure (leading to axial growth) more readily than on vdW bonded locations at the sides of the nanowire (which would result in lateral growth)[27].

When the islands become larger, after growth for longer times, it is possible to distinguish the orientation of these nuclei with respect to the substrate (Figs. SI3 and 4). The nuclei appear either flat and aligned with the substrate or with the growth direction at a small angle to the substrate. We attribute the randomly oriented growth direction to the amorphous nature[28] of the growth substrate, $SiO_2$/Si. The distribution of nanowire orientations at longer times may imply that the orientation formed early in the nucleation process. This leads to an overall picture of Mode 1 growth that involves adsorption of flux onto the surface, diffusion by short distances to form a relatively high density of nuclei with a distribution of orientations, and incorporation on the growing crystal surfaces to create well defined facets and an elongated morphology by preferential addition to the ends of the 1D atomic chains in the structure.

The Mode 2 nanowires show dramatically different morphology even for similar growth conditions. To probe the growth mechanism of Mode 2 nanowires, we show in Fig. 2 the effect of the growth time on the structure and dimensions. Figs. 2a-c are representative SEM images of Mode 2 nanowires synthesized for 5, 15 and 45 minutes while other growth parameters are kept the same; a 1-min growth is shown in Fig. SI5. Figs. 2d-f are tilted SEM images of the same growth times, displaying the morphology of the segmented nanowires from another perspective. We immediately notice that each nanowire segment does not grow flat on the (amorphous) $SiO_2$/Si substrate. Instead, typical growth is off the substrate with an angle of 8 ± 3 degrees as shown in Fig. 2e. A second distinctive feature of the experiments is the rapid appearance of the long segmented nanowires. Even after 1 min growth, structures as long as 50 μm are visible, composed of multiple segments that are narrow but still show height changes at their junctions. If each chain of segments is considered as a single nanostructure, then the density of nanostructures is around 100x lower (500-1000/cm$^2$) than for the Mode 1 nanowires ($10^5$/cm$^2$). We measure and plot the width and length of individual segments (as defined in Fig. SI6a-b) and the total length of the chained nanowires versus growth time in Figs. SI6c-e. We find that even though each chained nanowire may consist of a different number of segments, the segments, either within the same chain or in different chains, show only small variations in width, length, maximum height from the surface and tilt angle. The similarity in orientation suggests that each new segment is templated by the previous one.

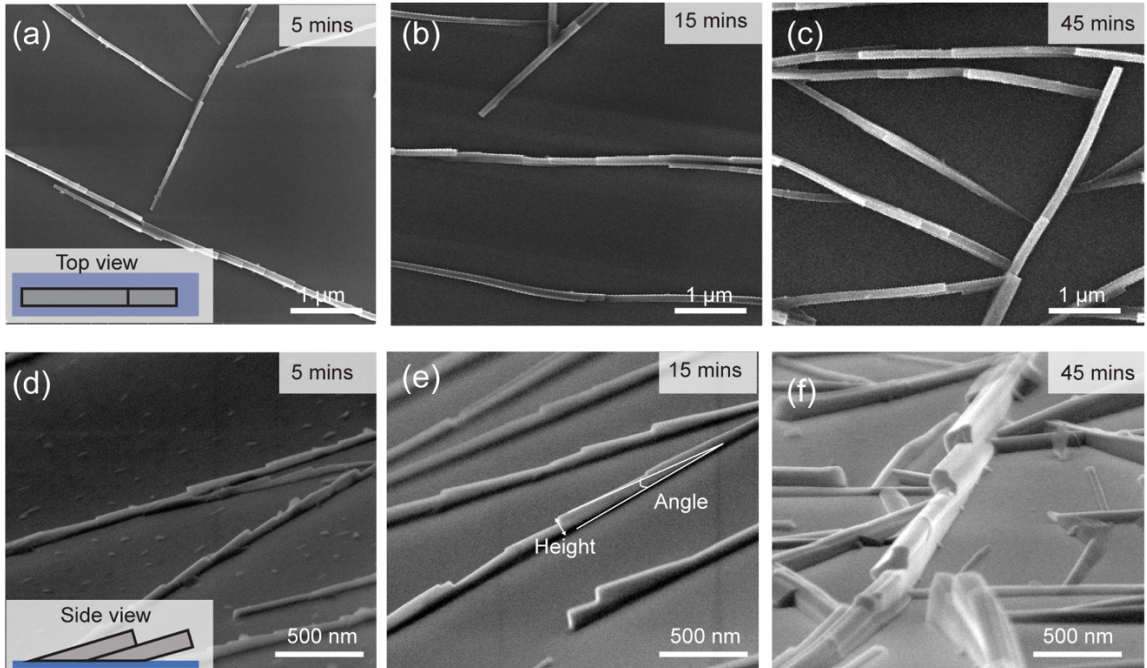

**Figure 2.** Time-dependent growths of Mode 2 NbS$_3$ nanowire chains. (a-c) Top-view SEM images of 5-, 15- and 45-minute growths. (d-f) Tilted SEM images of 5-, 15- and 45-minute growths. For an individual segment, the maximum height and the angle of the long axis off the substrate are indicated in panel e and Fig. SI6a.

As adsorbed atoms diffuse over the surface, it is unsurprising that their attachment is strongly favored at the location of an existing nanowire compared to initiating a new nanowire at another location on the surface: this is because a diffusing adatom is likely to find a low energy incorporation site on a crystal already present on the surface, compared to the larger barrier expected for forming a new critical size nucleus. We propose in Fig. 3 a model in which the unusual morphology seen in Mode 2 results from the combination of this favored incorporation and the misorientation between the initial nucleus and the substrate. Suppose that a faceted nucleus grows with the overall morphology described previously for Mode 1, where arriving atoms preferentially add to the end facets, but the nucleus formed with a small misorientation to the substrate. One end (the "high end") rises away from the substrate surface during growth, as shown in Fig. 3a. At the other end ("low end"), continued growth extends the long facets of the nanowire down towards the substrate, eventually eliminating the small ac-plane end facet. We suggest that this evolution in morphology of both ends has consequences for further growth of the structure and may increase the tendency to nucleate a poorly aligned section of new crystal at either end (Fig. 3b). At the high end, growth may elongate the nanowire to such an extent that it becomes difficult for adatoms supplied from the substrate to reach sites at the high end facet due to their finite diffusion distance, slowing growth at this end. At the low end, the adatoms do not encounter easy incorporation sites either once the low end facet (ac plane) disappears as the nanowire grows down to the surface. The favored incorporation site may now be at the contact line where the nanowire meets the surface, sites #1 and #2 in Fig. 3a, and will feel some influence of the substrate roughness or heterogeneous chemical nature. Continued growth at either site,

Fig. 3b, has a higher probability (compared to homogeneous growth) of incorporating a defect where the added material does not perfectly match the orientation of the initial segment.

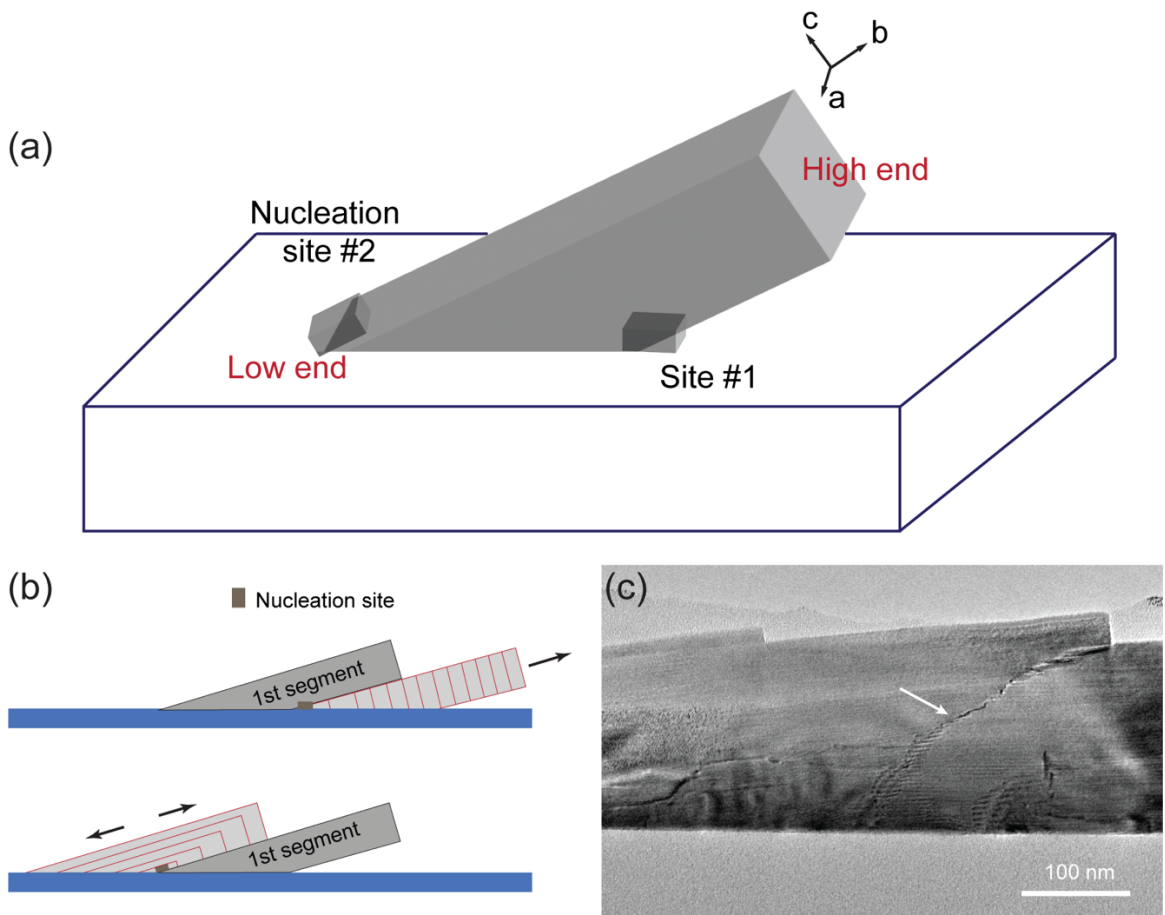

**Figure 3**. Proposed mechanism for Mode 2 nanowires. (a) A schematic showing the inclined angle between the first segment and the substrate. The length over which the segment grows off the substrate is exaggerated for clarity. Two possible nucleation sites (#1 and #2) are displayed. (b) The formation of a new segment templated by the exisiting segment. The black arrow indicates the growth direction in the two scenarios. Again the length over which the segment grows off the substrate is exaggerated. (c) Cross-sectional TEM image showing the interface (indicated by a white arrow) between segments in a Mode 2 nanowire, a curved interface assumed to trace the contact point during overgrowth of the two segments.

Our images of Mode 2 structures are consistent with this model. However, although nucleation site #2 appears plausible from images such as Fig. 4a (to be discussed below), it is not always clear in which direction a given chained nanowire had grown, and therefore whether nucleation site #1 or #2 was active. We therefore show in Fig. SI7 some morphologies exhibited by VLS-grown $NbS_3$ nanowires.[22] We used the same substrate as for the Mode 2 growth with the addition of Au-based eutectic droplets that offer preferential adsorption sites for the growth species. The VLS segments grow with a wide range of orientations with respect to the substrate (Fig. SI7a-b), but when the angle is small (below 10º) and the growth time is increased, we observe sawtooth surface structures as in Fig. SI7c-d. Since VLS growth dominates VS growth,[22] the VLS segment presumably grew first, meaning that the second segment appears to have nucleated at the contact line between the substrate and the misoriented initial segment. Although the presence of the catalyst changes the details of growth, this is an existence proof for nucleation at site #1 when the prior segment grows at a small angle off the substrate.

As growth continues, we observe that the Mode 2 structures evolve, not only increasing their length, width and height but also increasing the lengths of the segments and the height of the jumps where segments join (Fig. SI6). Since multi-segment NWs form even after short growth times (as in Figs. SI6d-e), and it is unlikely that the angles of segments change during and after growth, it appears that the changes in surface morphology are associated with the material that adds onto the surface of the initial, narrow segmented structure. The jumps in height have a similar appearance to the step bunches seen in conventional crystal growth. Growth by step flow is generally assumed to be unstable to bunching if there is an asymmetry in the attachment pathways of adatoms onto steps (Ehrlich-Schwoebel barrier[29]). The bunches themselves are expected to coalesce and coarsen. In models, the step bunch height and hence the separation between bunches increase sublinearly with growth time (for example[30] proportional to $t^{1/2}$). The separation between jumps in our experiments does increase with time, Fig. SI6e. The increase is faster than sublinear, but quantitative agreement with theory would be surprising as the growth fronts are highly three-dimensional, as seen in Fig. 3c, and the vertical facet is more complex in structure than a group of steps, Fig. SI2a. This change in the segment length with growth time suggests that the morphology seen after longer growth times arises from overgrowth of one segment to cover another and the visible height jumps do not represent the initial positions of the junctions between the segments. Further evidence that the position of the jump changes during growth comes from post-growth imaging using cross sectional TEM. Examples are shown in Fig. 1f and Fig. 3c. The shape of the junction, which does not appear as a straight line (as might be expected if a misaligned segment grows over a faceted segment), suggests competitive growth of both segments with the development of a thick overgrowth of the upper segment.

In short, we propose that to form the Mode 2 chained nanowires, initial nanowire segments grow as faceted elongated blocks. But misorientation between segments and the substrate surface leads to a complex morphology where additional nanowire segments nucleate at the contact line with the substrate, giving a probability of a small change in angle. The result is long curving nanowires composed of short straight segments with small angular misalignment. This model relies on several ingredients: a substrate that generates misoriented nuclei, with relatively low nucleation density, and a highly anisotropic growth process that spontaneously forms nuclei in the shape of elongated nanowires with incorporation strongly biased towards the ends. If the nucleation density is high, growth will take place as short nanowire segments competing for the growth flux, as seen in our Mode 1 growths (Fig.1a-c and Fig. SI8). On the other hand, at the far sides of the growth area (see inset of Fig. 1a), we speculate that the lower flux and therefore decreased nucleation density enables the segment ends to become the most likely nucleation sites and Mode 2 morphology to form.

### 2.3 Effect of the substrate on NbS$_3$ nanowire growth
A consequence of the role of the substrate in our model is that Mode 2 growth is only expected on substrates that produce *both* misoriented nuclei and low nucleation densities. We therefore tested this aspect of the growth hypothesis by carrying out growth on substrates expected to have different crystallinity, smoothness and diffusion properties.

We first grew $NbS_3$ nanowires on $SiO_2$/Si substrates on which thick (several to tens of layers) flakes of van der Waals materials had been exfoliated (Methods). Thick layers of van der Waals materials are expected to show low surface roughness when deposited on $SiO_2$, based on literature measurements[31]. Multilayer h-BN, graphene and CrSBr (a layered material with rectangular unit cell) were exfoliated onto $SiO_2$/Si substrates. These 2D materials exhibit different bonding and crystal symmetry, with graphene having the most isotropic bonding and CrSBr being anisotropic, with electronic structure that can be described as "one dimensional"[32]. We also grew the nanowires on a bulk crystalline substrate, c-oriented sapphire. These substrate characteristics influence the growth behavior of $NbS_3$ nanowires (Figure 4).

We find that $NbS_3$ nanowires on graphene and h-BN display the structural characteristics of Mode-1 nanowires, that is long, thin, and flat nanowires (Fig. SI10). Neither tilted nanowires nor Mode-2 nanowire chains were observed. The nanowires are instead parallel to the surface, although pointing in arbitrary orientations (Fig. SI10). The density of islands is different in each case. h-BN forms a high density of short nanostructures, indicating low diffusion distances. Graphene shows widely spaced and larger, longer nanowires, suggesting long diffusion distances[33]. These observations together confirm that the misorientation of the nuclei, and not just a low nucleation density, is an essential ingredient in the Mode 2 formation process.

$NbS_3$ nanowires grown on the non-planar parts of the 2D flakes show different morphologies that provide some insight into the nucleation process. The nanowires growing off the graphene edges toward the substrate (Figs. 4a-b) form parallel nanowire arrays, usually with a mix of both Mode 1 and Mode 2 morphologies. Since the graphene flakes were exfoliated using the Scotch tape method, they exhibit edges that are known to be dominantly zigzag or armchair structure[34,35]. It appears that the nanowires nucleating and growing off these graphene edges are either perpendicular to the edges or grow at approximately 60º to the edges (Figs. 4a-b). In h-BN flakes, the edge growth shows a somewhat different morphology that appears to depend on the flake thickness. Multilayer h-BN ~10-20 nm thick generates parallel arrays of Mode-2 chained nanowires growing across the $SiO_2$ surface, with some even bridging between flakes (Fig. 4c). The orientation distribution appears broader than that observed in graphene (Figs. 4a-b). We suggest that these chains may start their growth with an angle to the substrate, leading to the Mode 2 growth. For thin h-BN flakes (< 5 nm), $NbS_3$ nanowires grow along the circumference (Fig. 4d), creating 1D/2D vdW lateral heterostructures.

Finally, we demonstrate epitaxial growth of $NbS_3$ nanowires (Mode 1) on exfoliated CrSBr on $SiO_2$/Si (Fig. 4e), and on c-sapphire or (0001)-$Al_2O_3$, a 3D crystalline substrate (Fig. 4f). For the growth on CrSBr, the nanowires grow with preferred orientations with respect to the long edges of CrSBr flakes, Fig. 4e. Since CrSBr is known to cleave into flakes with long axis parallel to the CrSBr a-direction, the anisotropic structure of the CrSBr surface therefore appears to enable epitaxy with $[010]_{NbS_3}$ parallel to the [110] and [1-10] directions of the CrSBr. Some nanowires even nucleate on CrSBr and grow over to the uncovered $SiO_2$/Si area maintaining the epitaxial relationship. On the other hand, $NbS_3$ grown on c-sapphire shows an orientation relationship with three symmetry-equivalent growth orientations in which $[010]_{NbS_3}$ || $<11\text{-}20>_{Al_2O_3}$ as guided by the (0001) surface of the c-sapphire substrate (Fig. 4f). We note that on crystalline c-sapphire,

Mode 2 chained nanowires were not observed. This aligns with our suggestion above that a misaligned first segment is a necessary condition to initiate the sawtooth chain structure.

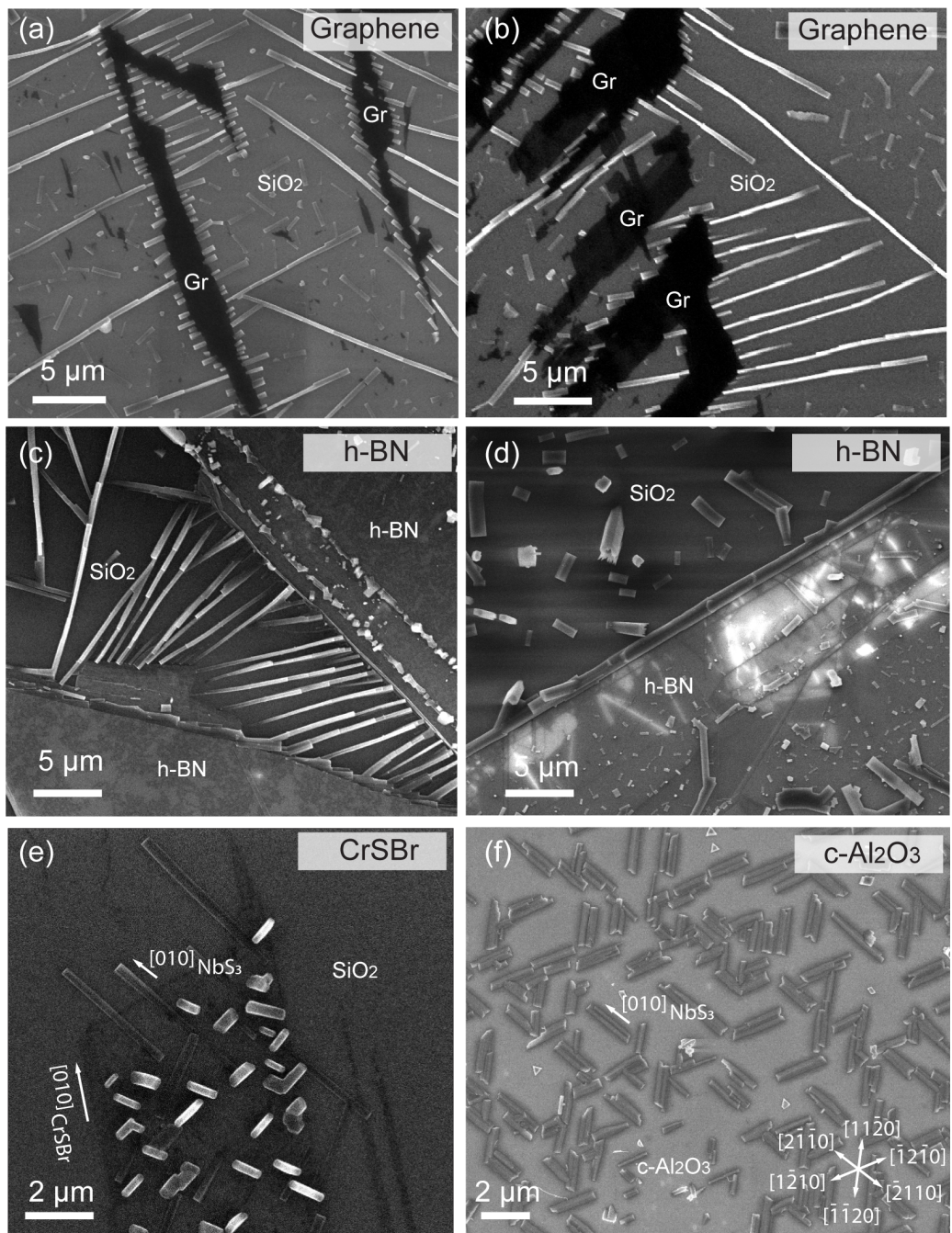

**Figure 4.** Directed assembly of 1D NbS$_3$ nanowires using 2D vdW flakes and 3D crystalline substrate. (a-b) SEM images of parallel arrays of NbS$_3$ nanowires growing from the edges of a graphene flake and bridging to adjacent flakes. (c-d) Growth and assembly of NbS$_3$ nanowires using h-BN. (e) SEM images of NbS$_3$ nanowires growing on the surface of CrSBr showing an epitaxial relationship. In (a-e), 2D materials were exfoliated and transferred onto SiO$_2$/Si substrates. (f) Epitaxial growth of NbS$_3$ nanowires on c-sapphire. The growth direction of NbS$_3$ wires and equivalent guided directions of c-sapphire susbtrate is shown.

We conclude that the use of 2D and 3D substrates with different crystallinity and symmetry can enable a range of growth morphologies. Given the lengths possible, substrates with spatial variations in crystallinity and topography are especially promising if we wish to combine the features of growth on amorphous, rough and crystalline surfaces. The substrate surface and, especially, the presence of nucleation sites such as the flake edges strongly affect $NbS_3$ nanowire nucleation, growth direction and morphology, i.e. Mode 1 or 2. This indicates the potential of 1D vdW patterned substrates for control of growth to realize dimensionally hybrid vdW heterostructures[1].

## 3. Conclusions

We have explored the chemical vapor deposition synthesis of nanowires in a transition metal trichalcogenide, $NbS_3$. The synthesis yields elongated nanocrystals, but the growth morphologies vary dramatically with conditions. We studied the impact of growth parameters on the growth mode, especially the role of the substrate structure, the flux rates and the growth time, to understand the origins of the morphology produced.

Measurements by SEM and STEM show that individual elongated and faceted nanostructures nucleate spontaneously under all circumstances studied. This stage of growth can be understood by considering the highly anisotropic crystal structure of $NbS_3$, where incorporation is likely to be preferred on the ends of the covalently bonded atomic chains rather than on the van der Waals surfaces, resulting in the formation of anisotropic structure (i.e., nanowires).

After nucleation, we find that the growth trajectory depends on the conditions. At the substrate location where the precursor fluxes are highest (near the center of the substrate facing the metal and salt precursor directly), we observe a high density of short nanowires. Towards the two ends of the growth area, further away from the precursor location, the nucleation density is lower, which we attribute to lower flux rates. The result is a strikingly different morphology - extremely long yet narrow nanowires composed of slightly misaligned segments connected end-to-end. To explain this unusual morphology, we develop a model depending on three key factors: nucleation density (controlled by diffusion length, among other factors), the degree of misorientation of nuclei with respect to the substrate (depending on substrate chemical nature and planarity), and the anisotropic nature of atom incorporation on the growing nuclei. Other growths using a catalyst-assisted method and growth on flat 2D (van der Waals flakes) and crystalline 3D (c-sapphire) substrates yield morphologies that are consistent with the predictions of the model.

We believe that our growth model for the chained nanowires may be more generally applicable, and we anticipate future experiments in which the growth mode can be evaluated in other anisotropic materials. Other transition metal trichalcogenides are an obvious next step, motivated by their range of interesting physical properties. An exciting consequence of the variability of the growth morphology is the opportunity to pattern the substrate to modulate the morphology produced. We expect that the configuration of self-assembled $NbS_3$ nanowires can be controlled by patterning through chemical or topographical means, and that we can exploit preferential nucleation sites to create crystalline structures that can bridge long distances on the substrate. For example, alternating amorphous and crystalline regions could modulate between Mode 1 and Mode 2 growth. This possibility of programable growth in combination with the remarkable length makes the chained nanowires a potential candidate for device applications such as electrical interconnects[36] or thermal dissipation components[37]. Finally, we note that the direct growth of 1D vdW nanowires on 2D vdW substrates that we have demonstrated here creates a clean, abrupt interface with presumed low contact resistance properties that will provide a platform to study proximity-induced effects in dimensionally hybrid heterostructures[38,39].

## 4. Experimental Section

**Substrate preparation**: 300-nm thermal $SiO_2$/Si and c-sapphire chips (Materials Supplies) were cleaned by acetone followed by isopropanol and deionized water. After drying, they underwent cleaning by oxygen plasma (150 mTorr) for 3-5 minutes. To create 2D substrates, few-layer graphene, h-BN and CrSBr were mechanically exfoliated directly onto the cleaned $SiO_2$/Si substrates using Scotch tape method. For observation on TEM grids, the 2D materials were suspended over a holey silicon nitride membrane using a cellulose acetate butyrate (CAB) polymer wet transfer method, followed by critical point drying to ensure clean surfaces.

**CVD growth**: All chemicals were from Sigma Aldrich. Nb powder (50-100 mg) was ground with NaCl (25 mg) before loading into an alumina boat. The second precursor, S powder of typically 100-200 mg, was placed in another boat upstream. A growth substrate was placed upside down on top of the Nb + NaCl boat and its temperature was set at 700-750°C. The temperature at the S boat location was set to 150-200°C. The S boat was kept at room temperature at the beginning of each growth and was pushed into this hot zone only after the Nb/NaCl/substrate boat reached the target temperature. High purity Ar (Airgas) with a flow rate of 100-150 sccm was used as the carrier gas. The duration of growth varied from 1 to 45 minutes. After the growth time, the carrier gas was increased to 300 sccm and the furnace cap was opened to help cool down the furnace quickly.

**Material characterization**: SEM images were obtained at 2-5 kV on an FEI Helios Nanolab 600 and Helios 5 UC. FIB samples were prepared using the 30 kV Ga ion beam on the same instrument. STEM characterization at 200 kV was obtained using a Themo Fisher Scientific Themis Z STEM operated at 200kV. Typical aberration-corrected STEM imaging conditions are 19 mrad semi-convergence angle (estimated sub-angstrom probe size) and 80 pA probe current. EDS was collected using a higher probe current of 120-200 pA with an acquisition time of 5-30 minutes. Elemental mapping and quantification used Thermo Fisher Velox software.

**Supporting Information**
Supporting Information is available from the Wiley Online Library or from the author.

**Acknowledgements:** We acknowledge funding from the U.S. Department of Energy, Office of Basic Energy Sciences, Division of Materials Sciences and Engineering under Award DE-SC0019336. K.R. acknowledges support from the Hugh Hampton Young Memorial Fellowship. This work was carried out in part through the use of MIT.nano and Virginia Tech's NCFL share faciliites.

**Conflict of Interest**
The authors declare no conflict of interest.

**Author contributions:** T.P. and F.M.R. conceived the project. A.N. and T.P. conducted the synthesis, SEM, STEM data collection and data analysis. K.R. contributed to the synthesis. T.P., M.A.F. and F.M.R. analyzed the data and wrote the manuscript with contributions from all authors.

**Data Availability Statement**
The data that support the findings of this study are available in the supplementary material of this article.

**Keywords**
Guided assembly, 1D van der Waals material, niobium trisulfide, nanowire chains

**References**

[1]  D. Jariwala, T. J. Marks, M. C. Hersam, *Nat. Mater. 2016 162* **2016**, *16*, 170.
[2]  A. A. Balandin, F. Kargar, T. T. Salguero, R. K. Lake, *Mater. Today* **2022**, *55*, 74.
[3]  J. O. Island, A. J. Molina-Mendoza, M. Barawi, R. Biele, E. Flores, J. M. Clamagirand, J. R. Ares, C. Sánchez, H. S. J. Van Der Zant, R. D'Agosta, I. J. Ferrer, A. Castellanos-Gomez, *2D Mater.* **2017**, *4*, 022003.
[4]  T. Pham, S. Oh, S. Stonemeyer, B. Shevitski, J. D. Cain, C. Song, P. Ercius, M. L. Cohen, A. Zettl, *Phys. Rev. Lett.* **2020**, *124*, 206403.
[5]  J. K. Qin, C. Wang, L. Zhen, L. J. Li, C. Y. Xu, Y. Chai, *Prog. Mater. Sci.* **2021**, *122*, 100856.
[6]  Y. Meng, W. Wang, J. C. Ho, *ACS Nano* **2022**, *16*, 13314.
[7]  R. V. Coleman, B. Giambattista, P. K. Hansma, A. Johnson, W. W. McNairy, C. G. Slough, *Adv. Phys.* **1988**, *37*, 559.
[8]  M. D. Randle, A. Lipatov, I. Mansaray, J. E. Han, A. Sinitskii, J. P. Bird, *Appl. Phys. Lett.* **2021**, *118*, 210502.
[9]  A. Patra, C. S. Rout, *RSC Adv.* **2020**, *10*, 36413.
[10] G. Grüner, A. Zettl, *Phys. Rep.* **1985**, *119*, 117.
[11] L. Yang, Y. Tao, Y. Zhu, M. Akter, K. Wang, Z. Pan, Y. Zhao, Q. Zhang, Y. Q. Xu, R. Chen, T. T. Xu, Y. Chen, Z. Mao, D. Li, *Nat. Nanotechnol. 2021 167* **2021**, *16*, 764.
[12] X. Chen, C. Zhu, B. Lei, W. Zhuo, W. Wang, J. Ma, X. Luo, Z. Xiang, X. Chen, *Phys. Rev. B* **2024**, *109*, 144513.
[13] S. Conejeros, B. Guster, P. Alemany, J.-P. Pouget, E. Canadell, *Chem. Mater.* **2021**, *33*, 5449.
[14] E. Zupanič, H. J. P van Midden, M. A. van Midden, S. Šturm, E. Tchernychova, V. Ya Pokrovskii, S. G. Zybtsev, V. F. Nasretdinova, S. V. Zaitsev-Zotov, W. T. Chen, W. Wu Pai, J. C. Bennett, A. Prodan, *Phys. Rev. B* **2018**, *98*, 174113.
[15] S. G. Zybtsev, V. Y. Pokrovskii, V. F. Nasretdinova, S. V. Zaitsev-Zotov, E. Zupanič, M. A. van Midden, W. W. Pai, *J. Alloys Compd.* **2021**, *854*, 157098.
[16] Z. Tian, C. Han, Y. Zhao, W. Dai, X. Lian, Y. Wang, Y. Zheng, Y. Shi, X. Pan, Z. Huang, H. Li, W. Chen, *Nat. Commun. 2021 121* **2021**, *12*, 1.
[17] J. M. Woods, Y. Jung, Y. Xie, W. Liu, Y. Liu, H. Wang, J. J. Cha, *ACS Nano* **2016**, *10*, 2004.
[18] L. Zhang, Q. Wang, L. Wang, L. Wang, J. Zhao, S. Li, *Phys. Rev. B* **2024**, *109*, 075306.
[19] T. Lv, X. Huang, W. Zhang, C. Deng, F. Chen, Y. Wang, J. Long, H. Gao, L. Deng, L. Ye, W. Xiong, *ACS Appl. Mater. Interfaces* **2022**, *14*, 48812.
[20] J. O. Island, M. Barawi, R. Biele, A. Almazán, J. M. Clamagirand, J. R. Ares, C. Sánchez, H. S. J. van der Zant, J. V. Álvarez, R. D'Agosta, I. J. Ferrer, A. Castellanos-Gomez, *Adv. Mater.* **2015**, *27*, 2595.
[21] R. S. Wagner, W. C. Ellis, *Appl. Phys. Lett.* **1964**, *4*, 89.
[22] T. Pham, K. Reidy, J. D. Thomsen, B. Wang, N. Deshmukh, M. A. Filler, F. M. Ross, *Adv. Mater.* **2024**, 2309360.
[23] M. A. Buckingham, B. Ward-O'Brien, W. Xiao, Y. Li, J. Qu, D. J. Lewis, *Chem. Commun.* **2022**, *58*, 8025.
[24] W. Han, K. Liu, S. Yang, F. Wang, J. Su, B. Jin, H. Li, T. Zhai, *Sci. China Chem.* **2019**, *62*, 1300.
[25] C. Xie, P. Yang, Y. Huan, F. Cui, Y. Zhang, *Dalton Trans.* **2020**, *49*, 10319.
[26] M. A. Bloodgood, P. Wei, E. Aytan, K. N. Bozhilov, A. A. Balandin, T. T. Salguero, *APL Mater.* **2017**, *6*, 026602.

[27] D. Lopez, Y. Zhou, D. L. M. Cordova, G. M. Milligan, K. S. Ogura, R. Wu, M. Q. Arguilla, *J. Am. Chem. Soc.* **2024**, *146*, 22863.

[28] V. O. Gridchin, K. P. Kotlyar, R. R. Reznik, L. N. Dvoretskaya, A. V. Parfen'eva, I. S. Mukhin, G. E. Cirlin, *Tech. Phys. Lett.* **2020**, *46*, 1080.

[29] M. H. Xie, S. Y. Leung, S. Y. Tong, *Surf. Sci.* **2002**, *515*, L459.

[30] F. Krzyżewski, M. Załuska-Kotur, A. Krasteva, H. Popova, V. Tonchev, *J. Cryst. Growth* **2017**, *474*, 135.

[31] D. Rhodes, S. H. Chae, R. Ribeiro-Palau, J. Hone, *Nat. Mater. 2019 186* **2019**, *18*, 541.

[32] J. Klein, T. Pham, J. D. Thomsen, J. B. Curtis, T. Denneulin, M. Lorke, M. Florian, A. Steinhoff, R. A. Wiscons, J. Luxa, Z. Sofer, F. Jahnke, P. Narang, F. M. Ross, *Nat. Commun.* **2022**, *13*, 1.

[33] J. D. Thomsen, K. Reidy, T. Pham, J. Klein, A. Osherov, R. Dana, F. M. Ross, *ACS Nano* **2022**, *16*, 10364.

[34] C. O. Girit, J. C. Meyer, R. Erni, M. D. Rossell, C. Kisielowski, L. Yang, C.-H. Park, M. F. Crommie, M. L. Cohen, S. G. Louie, A. Zettl, *Science* **2009**, *323*, 1705.

[35] K. Kim, S. Coh, C. Kisielowski, M. F. Crommie, S. G. Louie, M. L. Cohen, A. Zettl, *Nat. Commun.* **2013**, *4*, 2723.

[36] T. A. Empante, A. Martinez, M. Wurch, Y. Zhu, A. K. Geremew, K. Yamaguchi, M. Isarraraz, S. Rumyantsev, E. J. Reed, A. A. Balandin, L. Bartels, *Nano Lett.* **2019**, *19*, 4355.

[37] L. Yang, R. Prasher, D. Li, *J. Appl. Phys.* **2021**, *130*, 220901.

[38] R. Zhao, B. Grisafe, R. Krishna Ghosh, al -, W. Wen, C. Dang, J. Martincová, M. Otyepka, P. Lazar -, K. Szałowski, M. Milivojevi, D. Kochan, M. Gmitra, *2D Mater.* **2023**, *10*, 025013.

[39] N. Tilak, M. Altvater, S.-H. Hung, C.-J. Won, G. Li, T. Kaleem, S.-W. Cheong, C.-H. Chung, H.-T. Jeng, E. Y. Andrei, *Nat. Commun. 2024 151* **2024**, *15*, 1.